\documentstyle[12pt,epsf]{article}

\advance\voffset by -1.5cm
\advance\hoffset by -1.5cm
\def\be{\begin{equation}}
\def\ee{\end{equation}}

\def\pmb#1{\setbox0=\hbox{#1}
 \kern-.025em\copy0\kern-\wd0
 \kern.05em\copy0\kern-\wd0
 \kern-.025em\raise.0433em\box0 }

\def\I{{\cal I}}

\def\3{\ss}
\def\sq{\hbox{\rlap{$\sqcap$}$\sqcup$}}
\def\qed{\ifmmode\sq\else{\unskip\nobreak\hfil
\penalty50\hskip1em\null\nobreak\hfil\sq
\parfillskip=0pt\finalhyphendemerits=0\endgraf}\fi}
\def\half {\frac{1}{2}}

\def\bbbz {{\sf Z\!\!Z}}

\def\ss{\bf S}

\begin{document}

\thispagestyle{empty}
\def\thefootnote{\fnsymbol{footnote}}
\begin{flushright}
  hep-th/9901014\\
  CALT-68-2206 \\
  DAMTP-1998-170
 \end{flushright}
\vskip 0.5cm

\begin{center}\LARGE
{\bf Non-BPS States in Heterotic -- Type IIA Duality}
\end{center}
\vskip 1.0cm
\begin{center}
{\large  Oren Bergman\footnote{E-mail  address: 
{\tt bergman@theory.caltech.edu}}}

\vskip 0.5 cm
{\it Department of Physics\\
California Institute of Technology\\
Pasadena, CA 91125}

\vskip 1.0 cm
{\large  Matthias R. Gaberdiel\footnote{E-mail  address: 
{\tt M.R.Gaberdiel@damtp.cam.ac.uk}}}

\vskip 0.5 cm
{\it Department of Applied Mathematics and Theoretical Physics \\
University of Cambridge, Silver Street, \\
Cambridge CB3 9EW, U.K.}
\end{center}

\vskip 1.0cm

\begin{center}
January 1999
\end{center}

\vskip 1.0cm

\begin{abstract}
The relation between some perturbative non-BPS states of the heterotic
theory on $T^4$ and non-perturbative non-BPS states of the orbifold 
limit of type IIA on K3 is exhibited. 
The relevant states include a non-BPS D-string, 
and a non-BPS bound state of BPS D-particles (`D-molecule').
The domains of stability of these states in the two theories are
determined and compared.
\end{abstract}

\vskip 1.0cm 
\begin{center}
\end{center}

\vfill
\setcounter{footnote}{0}
\def\thefootnote{\arabic{footnote}}
\newpage

\renewcommand{\theequation}{\thesection.\arabic
{equation}}

\section{Introduction}
\setcounter{equation}{0}

It has become apparent recently that the duality symmetries of string
theory give rise to predictions about states that are not necessarily
BPS \cite{Sen1}. In particular, it was demonstrated by Sen in
\cite{Sen2,Sen3} that perturbative non-BPS states of one theory, that
are stable due to the fact that they are the lightest states carrying
a given set of charges, can sometimes be identified in the dual theory
as bound states of BPS D-strings and anti-D-strings. Alternatively,
these states could be described as novel non-BPS D-particles
\cite{BG2,Sen4} that are easily constructed using the boundary state 
approach \cite{PolCai,CLNY,Li,CallanKlebanov,BG1,Billo}. These, and
perhaps other non-BPS D-branes can also be naturally understood in
terms of K-theory \cite{Witten2}, where different states that can
decay into one another lie in the same equivalence class. 

So far only two cases have been studied in detail. In one of them, the
perturbative non-BPS state in question is a massive state in the
ten-dimensional $SO(32)$ heterotic string that transforms in the
spinor representation of $SO(32)$. The dual type I state is then a
$\bbbz_2$-valued non-BPS D-particle \cite{Sen4,Witten2}.  The other
case involves the orientifold theory IIB$/\Omega\I_4$, where $\I_4$ is
the inversion of four spatial directions. The relevant perturbative
non-BPS state in this case is the ground state of the string beginning
on a D5-brane and ending on its image, and the dual state is a non-BPS
D-particle in the orbifold of type IIB by $(-1)^{F_L} \I_4$
\cite{BG2}.

If we compactify the latter theory on a $4$-torus, the orbifold is
related by T-duality to type IIA on $T^4/\I_4$, which in turn is the
orbifold limit of a K3 surface. On the other hand, IIA on K3 is also
related non-perturbatively to the heterotic string on $T^4$, and it
should therefore be possible to identify suitable perturbative non-BPS
states of the heterotic string on $T^4$ with brane states in IIA on
K3. This is all the more interesting since the (BPS) spectrum of the
heterotic string is very well understood, and therefore detailed
comparisons can be made.

In this paper we analyse two classes of perturbative non-BPS states in
the heterotic theory on $T^4$, and relate them to non-BPS 
states in the $T^4/\I_4$ orbifold of IIA. The relevant states
are a non-BPS D-string (that is T-dual to the non-BPS
D-particle of the IIB orbifold theory), and a non-BPS `D-molecule'.
Both can be understood as non-BPS bound states of BPS D-branes;
in the first case tachyon condensation takes place, whereas in the
second the bound state is due to an ordinary attractive force.
In each case, we also analyse the stability of these non-BPS states as
a function of the moduli of the theory.  Since their masses are not 
protected against quantum corrections however, this analysis only
holds at weak coupling in either theory, and therefore we do not
expect the masses and regions of stability to be related by the
duality.  Nevertheless, we find non-vanishing regions of stability for
both types of states in both the heterotic string and the type IIA
string descriptions. Furthermore, for the non-BPS D-string these
regions are closely related by the duality transformation.

The orbifold limit of K3 is a rather special point in the moduli space
of the theory, and it is therefore interesting to understand how the
various states can be understood for a smooth K3. To this end we also
discuss how the non-BPS states can be understood in terms of wrapped
membranes.
\medskip

The paper is organised as follows: in section~2 the duality map
between the heterotic string on $T^4$ and type IIA at the orbifold
point of K3 is established. This is then tested by comparing the
masses of certain BPS states of the two theories in section~3. In
section~4 the non-BPS states are analysed in both theories. We
conclude and raise some open problems in section~5.
\bigskip

While this paper was being prepared we obtained the preprint
\cite{Sen5} in which the non-BPS D-string of the type IIA orbifold, as
well as its stability and interpretation in terms of wrapped
membranes, is also discussed.

\section{Heterotic -- IIA duality in the orbifold limit}
\setcounter{equation}{0}

Let us recall the precise relation between type IIA at the orbifold
point of K3 and the heterotic string on $T^4$; the following
discussion follows closely \cite{Polbook}. Denote the compact
coordinates by $x^i$, where $i=1,2,3,4$, and the corresponding radii
in the heterotic string theory by $R_{hi}$. The sequence of dualities
relating the two theories is given by   
\be
 \mbox{het}\;\; T^4 \stackrel{S}{\longrightarrow}
 \mbox{I}\;\; T^4 \stackrel{T^4}{\longrightarrow}
 \mbox{IIB}\;\; T^4/\bbbz_2' \stackrel{S}{\longrightarrow}
 \mbox{IIB}\;\; T^4/\bbbz_2'' \stackrel{T}{\longrightarrow}
 \mbox{IIA}\;\; T^4/\bbbz_2 \,,
\label{sequence}
\ee
where the various $\bbbz_2$ groups are 
\be
 \bbbz_2' = (1,\Omega\I_4) \quad
 \bbbz_2'' = (1,(-1)^{F_L}\I_4) \quad
 \bbbz_2 = (1,\I_4)\,.
\ee
Here $\I_4$ reflects all four compact directions, $\Omega$ reverses
world-sheet parity, and $F_L$ is the left-moving part of the spacetime
fermion number. The first step is ten-dimensional S-duality between 
the ($SO(32)$) heterotic string and the type I string \cite{PolWit},
which relates the (ten-dimensional) couplings and radii
as\footnote{Numerical factors are omitted until the last step.} 
\be
g_I \propto g_h^{-1} \qquad \qquad 
R_{Ij} \propto g_h^{-1/2} R_{hj} \,.
\ee
The second step consists of four T-duality transformations on the
four circles, resulting in the new parameters
\be
\begin{array}{lclcl}
g' & = & V_I^{-1} g_I & \propto & V_h^{-1} g_h \\
R'_j & = & R_{Ij}^{-1} & \propto & g_h^{1/2} R_{hj}^{-1} \,,
\end{array}
\ee
where $V_I=\prod_j R_{Ij}$ and $V_h= \prod_j R_{hj}$ denote the
volumes (divided by $(2\pi)^4$) of the $T^4$ in the type I and
heterotic strings, respectively. This theory has $16$ orientifold fixed
points. In order for the dilaton to be a constant, the RR charges have
to be cancelled locally, {\it i.e.} one pair of D5-branes has to be
situated at each orientifold 5-plane. In terms of the original
heterotic theory, this means that suitable Wilson lines must be
switched on to break  $SO(32)$ (or $E_8\times E_8$) to $U(1)^{16}$;
this will be further discussed below. The third step is S-duality of
type IIB. The new parameters are given by
\be
\begin{array}{lclcl}
g'' & = & g'^{-1} & \propto & V_h g_h^{-1} \\
R''_j & = & g'^{-1/2} R'_j & \propto & V_h^{1/2} R_{hj}^{-1} \,.
\end{array}
\ee
Finally, the fourth step is T-duality along one of the compact
directions, say $x^4$. The resulting theory is type IIA on a K3 in the
orbifold limit. The coupling constants and radii are given by 
\be
\label{relation}
\begin{array}{lclcl}
g_A & = & g'' (R''_4)^{-1} & = & g_h^{-1} R_{h4} V_h^{1/2} \\
R_{Aj} & = & R''_j & = & 2 V_h^{1/2} R_{hj}^{-1} \qquad 
\mbox{for $j\ne 4$} \\
R_{A4} & = & (R''_4)^{-1} & = & 2^{-1} V_h^{-1/2} R_{h4}\,,
\end{array}
\ee
where we have now included the numerical factors (that will be shown
below to reproduce the correct masses for the BPS-states).\footnote{In
our conventions $\alpha'_h=1/2$, $\alpha'_A=1$.} In addition, the
metrics in the low energy effective theories are related as
\cite{Witten1}   
\be
  G^{A}_{\mu\nu} = V_h g_h^{-2} G^{h}_{\mu\nu} \,.
\label{metric}
\ee

The appropriate Wilson lines in the heterotic theory on $T^4$ can be 
determined in analogy with the duality between the heterotic
string on $S^1$ and type IIA on $S^1/\Omega\I_1$ (type IA).
A constant dilaton background for the latter requires the
Wilson line $A=((\half)^8,0^8)$ in the former \cite{ks,lowe,bgl},
resulting in the gauge group $SO(16)\times SO(16)$.  
The sixteen entries in the  Wilson line describe the 
positions of the D8-branes along the interval in type IA.
This suggests that the four Wilson lines in our case should be
\begin{eqnarray}
A^1 & = & {\displaystyle \left(\left(\half\right)^8, 0^8 \right)}
  \nonumber \\
A^2 & = & {\displaystyle \left(\left(\half\right)^4,0^4,
           \left(\half\right)^4,0^4 \right)} \nonumber \\ 
A^3 & = & {\displaystyle \left(\left(\half\right)^2,0^2,
           \left(\half\right)^2,0^2,\left(\half\right)^2,0^2,
           \left(\half\right)^2,0^2 \right) }\nonumber \\
A^4 & = & {\displaystyle \left(\half,0,\half,0,\half,0,\half,0,
                \half,0,\half,0,\half,0,\half,0 \right)} \,,
\end{eqnarray}
so that there is precisely one pair of D-branes at each of the sixteen
orientifold planes. Indeed, this configuration of Wilson lines 
breaks the gauge group $SO(32)$ to $SO(2)^{16}\sim U(1)^{16}$, 
and there are no other massless gauge particles that are charged under
the Cartan subalgebra of $SO(32)$. To see this, recall that the momenta
of the compactified heterotic string are given as \cite{Ginsparg} 
\be
\begin{array}{lclcl}
{\bf P}_L & = & (P_L,p_L) & = & {\displaystyle
\left( V_K + A_K^i w_i\; ,\; {p^i \over 2 R_i}+ w^i R_i\right)}\\[10pt]
{\bf P}_R & = & \phantom{(P_R,} p_R & = & {\displaystyle
\left( {p^i \over 2 R_i} - w^i R_i  \right)\,,}
\end{array}
\ee
where $p^i$ is the physical momentum in the compact directions 
\be
 p^i = n^i + B^{ij} w_j - V^K A_K^i - {1\over 2}A_K^i A_K^j w_j \,,
\ee
$w_i,n_i\in\bbbz$ are elements of the compactification lattice
$\Gamma^{4,4}$, and $V^K$ is an element of the internal lattice
$\Gamma^{16}$. For a given momentum $({\bf P}_L,{\bf P}_R)$, a
physical state can exist provided the level matching condition    
\be
\label{level}
\half {\bf P}_L^2 + N_L - 1= \half {\bf P}_R^2 + N_R - c_R
\ee
is satisfied, where $N_L$ and $N_R$ are the left- and right-moving 
excitation numbers, and $c_R=1/2$ ($c_R=0$) for the right-moving NS
(R) sector. The state is BPS if $N_R=c_R$ \cite{dh}, and its mass is
given by  
\be
 {1\over 4}m_h^2 = \left(\half {\bf P}_L^2 + N_L - 1\right) +
                 \left(\half {\bf P}_R^2 + N_R - c_R\right) =
       {\bf P}_R^2 + 2 ( N_R - c_R)\,.
\label{heteroticmass}
\ee

The massless states of the gravity multiplet and the Cartan subalgebra
have $N_L=1$ and ${\bf P}_L={\bf P}_R=0$.  Additional massless gauge
bosons would have to have $N_L=0$, and therefore ${\bf P}_L^2 =2$.  If
$w_i=0$ for all $i$, this requires $V^2=2$ and $p_i=0$.  The possible
choices for $V$ are then simply the roots of $SO(32)$, and it is easy
to see that for each root at least one of the inner products 
$V^K A_K^i$ is half-integer; thus $p^i\in\bbbz + 1/2$ cannot vanish,
and the state is massive.  On the other hand, if $w_i\ne 0$ for at
least one $i$, the above requires $(V+Aw)^2<2$, and it follows that 
$V+Aw=0$, {\it i.e.} that the massless gauge particle is not charged
under the Cartan subalgebra of $SO(32)$.

\section{BPS states}
\setcounter{equation}{0}

In order to test the above identification further, it is useful to
relate some of the perturbative BPS states of the heterotic string to
D-brane states in IIA on $T^4/\bbbz_2$, and to compare their masses. 
Let us start with the simplest case -- a bulk D-particle.
This state is charged only under the bulk $U(1)$ corresponding
to the ten-dimensional RR one-form $C_{RR}^{(1)}$.
It can be described by the boundary state in IIA
\be
\label{D00}
|D0;\epsilon_1> = {1\over\sqrt{2}}\Big(
|U0>_{NSNS} + \epsilon_1 |U0>_{RR}\Big) \,,
\ee
where the two components are defined in the standard way \cite{BG1},
and lie in the untwisted NSNS and RR sectors, respectively. Here
$\epsilon_1=\pm 1$ differentiates a D-particle from an
anti-D-particle. For a suitable normalisation of the
two components the open-closed consistency condition is
satisfied \cite{Pol1,BG1}, and the spectrum of open strings 
beginning on one D-particle and ending on another is given by 
\be
[NS - R] \; {1\over 2} \Big(1+\epsilon_1\epsilon_1'(-1)^F\Big) \,.
\ee
The corresponding state in the heterotic string has trivial winding
($w_i=0$) and momentum ($V=0$, $p^i=0$), except for
$p_4=\epsilon_1$. Level matching then requires that $N_L=1$, and
therefore the state is really a Kaluza-Klein excitation of either the
gravity multiplet or one of the vector multiplets in the Cartan
subalgebra. Its mass is given by (\ref{heteroticmass}) 
\be
 m_h(D0) = {1\over R_{h4}} \,.
\ee
The corresponding mass in type IIA can be found using (\ref{relation})
and (\ref{metric}), and turns out to be  
\be
 m_A(D0) = V_h^{-1/2} g_h m_h(D0) = {1\over g_A} \,.
\ee
This is in complete agreement with the mass of a D-particle. 

Next consider a D-particle which is stuck at one of the fixed
planes. Both its mass and bulk RR charge are half of those of 
the bulk D-particle (since prior to the projection it corresponds to 
a single D-particle, whereas the bulk D-particle corresponds to
two D-particles); it is therefore called a `fractional' D-particle
\cite{DM}. It also carries unit charge with respect to the twisted RR
$U(1)$ at the fixed plane. The corresponding boundary state is of the
form  
\be
\label{D0}
|D0_f;\epsilon_1,\epsilon_2> = {1 \over 2}\bigg[\Big(|U0>_{NSNS} +
\epsilon_1|U0>_{RR}\Big) + \epsilon_2\Big(|T0>_{NSNS} +
\epsilon_1|T0>_{RR}\Big)\bigg] \,, 
\ee 
where $|U0>_{NSNS}$ and $|U0>_{RR}$ are the same states that appeared
in (\ref{D00}), and $|T0>_{NSNS}$ and $|T0>_{RR}$ lie in the twisted
NSNS and twisted RR sectors, respectively. Here $\epsilon_1=\pm 1$ and
$\epsilon_1\epsilon_2=\pm 1$ determine the sign of the bulk and the
twisted charges of the state, respectively. Using standard techniques
\cite{BG1,Sen2,BG2} it is easy to see that each of the components is
invariant under the GSO and orbifold projections, and that for a
suitable normalisation of the twisted components the open-closed
consistency condition is again satisfied. Indeed, the spectrum of open
strings beginning on one fractional D-particle and ending on another
is given by 
\be 
[NS - R] \; {1\over 4} \Big(1+\epsilon_1\epsilon_1'(-1)^F\Big)
\Big(1+\epsilon_2\epsilon_2'\I_4\Big)\,.  
\label{frac_D0_open}
\ee 
In the blow up of the orbifold to a smooth K3, the fractional
D-particle corresponds to a D2-brane which wraps a supersymmetric
cycle \cite{Douglas}. In the orbifold limit the area of this cycle
vanishes, but the corresponding state is not massless, since the
two-form field $B^{(2)}$ has a non-vanishing integral around the cycle
\cite{Aspinwall}. In fact $B=1/2$, and the resulting state carries one
unit of twisted charge coming from the membrane itself, and one half
unit of bulk charge coming from the D2-brane world-volume action term
$\int d^3\sigma \, C_{RR}^{(1)}\wedge (F^{(2)}+B^{(2)})$.  At each
fixed point there are four such states, corresponding to the two
possible orientations of the membrane, and the possibility of 
having $F=0$ or $F=\pm 1$ (as $F$ must be integral, the state always has
a non-vanishing bulk charge). These are the four possible fractional
D-particles of (\ref{D0}). Since there are sixteen orbifold fixed
planes, there are a total of $64$ such states.

In the heterotic string these correspond to states with internal
weight vectors of the form
\be
 V=\pm (0^{2n},1,\pm 1, 0^{14-2n}) \qquad (n=1,\ldots,8)\,,
\label{D0weights}
\ee
and vanishing winding and internal momentum, except for $p_4=\pm 1/2$.
The sixteen twisted $U(1)$ charges in the IIA picture correspond to
symmetric and anti-symmetric combinations of the $(2n+1)$'st and
$(2n+2)$'nd Cartan $U(1)$ charges in the heterotic picture. It follows
from the heterotic mass formula (\ref{heteroticmass}) that the mass of
these states is  
\be
 m_h(D0_f) = {1\over 2R_{h4}} \,.
\ee
As before, this corresponds to the mass
\be
 m_A(D0_f) = V_h^{-1/2} \, g_h \, m_h(D0_f) 
= {1\over 2g_A} \,,
\label{fracD0}
\ee
in the orbifold of type IIA, and is thus in complete agreement with
the mass of a fractional D-particle. 

Additional BPS states are obtained by wrapping D2-branes around
non-vanishing supersymmetric 2-cycles, and by wrapping D4-branes
around the entire compact space. 
One can compute the mass of each of these states, and thus find the
corresponding state in the heterotic string. Let us briefly summarise
the results: 
\begin{list}{(\roman{enumi})}{\usecounter{enumi}}
\item A D2-brane that wraps the cycle $(x^i,x^j)$ where $i\ne j$ and 
$i,j\in\{1,2,3\}$ has mass $m_A = R_{Ai} R_{Aj} / (2g_A)$; in heterotic
units this corresponds to $m_h=2 R_{hk}$, where $k\in\{1,2,3\}$ is not
equal to either $i$ or $j$. The corresponding heterotic state has 
$w_k=\pm 1$, $p^l=0$, $(V \pm A_k)^2=2$, and $N_L=0$.
\item A D2-brane that wraps the cycle $(x^i,x^4)$, where $i$ is either 
$1,2$ or $3$, has mass $m_A=R_{Ai} R_{A4} / (2g_A)$; in heterotic units
this corresponds to $m_h=1/(2R_{hi})$. The corresponding heterotic
state therefore has $p^i=\pm 1/2$, $w^j=0$, $V^2=2$, and $N_L=0$.
\item A D4-brane wrapping the entire compact space has mass
$m_A = \prod_i R_{Ai} / (2g_A)$; in heterotic units this corresponds to 
$m_h= 2 R_{h4}$.  The corresponding heterotic state therefore has
$w_4=\pm 1$, $p^l=0$, $(V \pm A_4)^2=2$, and $N_L=0$.
\end{list}

\section{Non-BPS states}
\setcounter{equation}{0}

\noindent The heterotic string also contains non-BPS states that are 
stable in certain domains of the moduli space. One should therefore
expect that these states can also be seen in the dual type IIA theory,
and that they correspond to non-BPS branes. Of course, since non-BPS
states are not protected by supersymmetry against quantum corrections
to their mass, the analysis below will only hold for $g_h\ll 1$ and
$g_A\ll 1$ in the heterotic and type IIA theory, respectively.

\subsection{Non-BPS D-string}

\noindent The simplest examples of this kind are the heterotic 
states with vanishing winding and momenta ($w_i=p_i=0$), and weight
vectors given by 
\be
\label{intern}
\begin{array}{lcl}
V &=& \left(0^m,\pm 2,0^{15-m}\right) \\
V' &=& \left(0^{2m},\pm 1,\pm 1, 0^{2n},\pm 1,\pm 1,0^{12-2n-2m}
    \right)\,.
\end{array}
\ee
The results of the previous section indicate that these states are 
charged under precisely two $U(1)$'s associated with two fixed points
in IIA, and are uncharged with respect to any of the other
$U(1)$'s. There are four states for each pair of $U(1)$'s, carrying
$\pm 1$ charges with respect to the two $U(1)$'s. In all cases
$V^2=4$, and we must choose $N_R=c_R+1$ to satisfy
level-matching. These states are therefore {\em not} BPS, and
transform in long multiplets of the $D=6$ ${\cal N}=(1,1)$ 
supersymmetry algebra. Their mass is given by
\be
 m_h = 2\sqrt{2}\,,
\label{het_nonbps_mass}
\ee
as follows from (\ref{heteroticmass}); in particular, the mass is
independent of the radii. 

On the other hand, these states carry the same charges as two BPS
states of the form discussed in the previous section (where the charge
with respect to the spacetime $U(1)$'s is chosen to be opposite for
the two states), and they might therefore decay into them. Whether or
not the decay occurs depends on the values of the radii, since the
masses of the BPS states depend on them. In particular, the first
state in (\ref{intern}) carries the same charges as the two BPS 
states with $p_4=\pm 1/2$, and weight vectors of the form
\begin{eqnarray}
 V_1 &=&  \pm (0^{2n},1,1, 0^{14-2n}) \nonumber\\
 V_2 &=&  \pm (0^{2n},1,-1, 0^{14-2n}) \,,
\label{het_decay1}
\end{eqnarray}
where $n=[m/2]$. The mass of each of these states is $1/(2R_{h4})$,
and the decay is therefore energetically forbidden when  
\be 
R_{h4} < {1\over 2\sqrt{2}} \,.  
\label{het_stab1}
\ee
More generally, the above non-BPS state has the same charges as two
BPS states with $w_i=0$, and internal weight vectors
\begin{eqnarray}
V_1 & = & \pm \left(0^m,1,0^{k},1,0^{14-m-k}\right) \nonumber \\
V_2 & = & \pm \left(0^m,1,0^{k},-1,0^{14-m-k}\right) \,,
\label{het_decay2}
\end{eqnarray}
where again the non-vanishing internal momenta are chosen to be
opposite for the two states. The lightest states of this form 
have a single non-vanishing momentum, $p_i=\pm 1/2$ for one of
$i=1,2,3,4$, and their mass is $1/(2R_{hi})$. Provided that 
\be
 R_{hi} < {1\over 2\sqrt{2}} \qquad i=1,2,3,4 \,,
\label{het_stab2}
\ee
the non-BPS state cannot decay into any of these pairs of BPS states,
and it should therefore be stable. 
Similar statements also hold for the 
non-BPS states of the second kind in (\ref{intern}).
\medskip

We should therefore expect that the IIA theory possesses a non-BPS
D-brane that has the appropriate charges and multiplicities.
This state is easily constructed: it is the non-BPS D-string of type
IIA, whose boundary state is given as   
\be
\label{D1p}
|D1_{nonbps};\theta,\epsilon> = 
  {1\over\sqrt{2}}\bigg[|U1;\theta>_{NSNS} 
+ {\epsilon\over\sqrt{2}} \left( |T1;1>_{RR} + e^{i\theta} |T2;2>_{RR}
  \right)\bigg]  \,,
\ee
where we have used the notation of Sen \cite{Sen2}.\footnote{This
state has also been independently constructed by Sen \cite{Sen5}.}
Here $\theta$ is the value of the Wilson line on the D-string,
which must be $0$ or $\pi$ in the orbifold, and $\epsilon=\pm 1$. 
The two states in the twisted RR sector are localised at either end of
the D-string (so that the D-string stretches between two orbifold
points). Using the standard techniques \cite{BG1,Sen2}, one can easily
check that each of the boundary components is invariant under the GSO
and orbifold projections, and, for a suitable normalisation of the 
different components, the open-closed consistency condition is
satisfied. The spectrum of open strings beginning and ending on the
same D-string is obtained as usual by computing the cylinder amplitude
with the above boundary state, and the result is  
\be
[NS - R ] \; {1 \over 4} \left( 1+ (-1)^F \I_4 \right) 
          \left( 1 + (-1)^F \I'_4 \right) \,,
\label{open_D1}
\ee
where $\I'_4$ is the same as $\I_4$, except that it acts on
$x^4$ as $x^4\rightarrow 2\pi R_{A4} - x^4$. For each pair of orbifold
points there are four D-strings, which are charged only under the two
twisted sector $U(1)$'s associated to the two orbifold points. These
charges are of the same magnitude as those of the  fractional
D-particles, since the ground state of $|T1>_{RR}$ is the same as that
of $|T0>_{RR}$ in (\ref{D0}). Furthermore, it follows from
(\ref{open_D1}) that the D-strings have sixteen (rather than eight)
fermionic zero modes, and therefore transform in long multiplets of
the $D=6$, ${\cal N}=(1,1)$ supersymmetry algebra. These states
therefore have exactly the correct properties to correspond to the
above non-BPS states of the heterotic theory.  

We should not, however, expect that the corresponding masses are
related by the duality map, since for non-BPS states the masses are
not protected from quantum corrections. Let us consider for example
the case where the D-string is suspended between two orbifold
points that are separated along $x^4$ (Fig.~1a). Its 
classical mass is given by 
\be
 m_A(D1_{nonbps}) = 
{R_{A4}\over \sqrt{2} g_A} \,. 
\label{D1mass}
\ee
The numerical factor can be determined by comparing the boundary state
of the non-BPS D-string (\ref{D1p}) to that of a BPS D-string between
two fixed planes in the (T-dual) type IIB orbifold (eq.~(3.16) of
\cite{Sen2}). The units of the two orbifold theories are simply
related by replacing $g_A$ with $g_B$, and since the coefficient of
the untwisted NSNS component is greater by a factor of $\sqrt{2}$ 
for the non-BPS D-string, its mass is given by (\ref{D1mass}).
In heterotic units, this mass is $\propto 1/V_h$, and therefore does
not agree with (\ref{het_nonbps_mass}).  

The open string NS sector in (\ref{open_D1}) contains a
tachyon. However, since the tachyon is $(-1)^F$-odd, and since $\I_4$
reverses the sign of the momentum along the D-string, the
zero-momentum component of the tachyon field on the D-string is
projected out. Furthermore, since $\I_4\I'_4$ acts as 
$x^4\rightarrow x^4 - 2\pi R_{A4}$, the half-odd-integer momentum 
components are also removed, leaving a lowest mode of unit momentum.
As a consequence, the mass of the tachyon is shifted to 
\be
 m^2_T =  - \half + {1\over R_{A4}^2} \,.
\ee
For $R_{A4}<\sqrt{2}$ the tachyon is actually massive, and thus
attains its vacuum value at the origin. On the other hand, for 
$R_{A4}>\sqrt{2}$ the tachyon has a non-zero vacuum expectation
value, and the lowest momentum mode describes a kink-anti-kink
configuration along the D-string, in which the tachyon field vanishes
at the two endpoints of the D-string, and approaches its vacuum value
in-between. For $R_{A4}>\sqrt{2}$ the state is therefore more 
appropriately described as a pair of fractional BPS D-particles
located at either fixed point, and carrying opposite bulk charges
(Fig.~1b). Alternatively, the ground state of the NS sector open
string between the above two fractional BPS D-particles has a mass 
\be
m^2 = -\half + \left(\pi R_{A4} T_0\right)^2
  = -\half + \left({R_{A4}\over 2}\right)^2\,,
\ee
and so becomes tachyonic for $R_{A4}<\sqrt{2}$, indicating an
instability to decay into the non-BPS D-string. 
The D-string can therefore be thought of as a bound state
of two fractional BPS D-particles located at different fixed
planes. This is also confirmed by the fact that the classical mass of
the D-string (\ref{D1mass}) is smaller than that of two fractional 
D-particles (\ref{fracD0}) when 
\be 
 R_{A4}< \sqrt{2} \,,
\label{IIA_stab1}
\ee
and thus the D-string is stable against decay into two fractional
D-particles in this regime. In terms of the heterotic string, this
decay channel corresponds to (\ref{het_decay1}). The regimes of
stability of the non-BPS state in the two dual theories,
(\ref{het_stab1}) and  (\ref{IIA_stab1}), are qualitatively the same, 
given the duality relation (\ref{relation}). 
\begin{figure}[htb]
\epsfxsize=5 in
\centerline{\epsffile{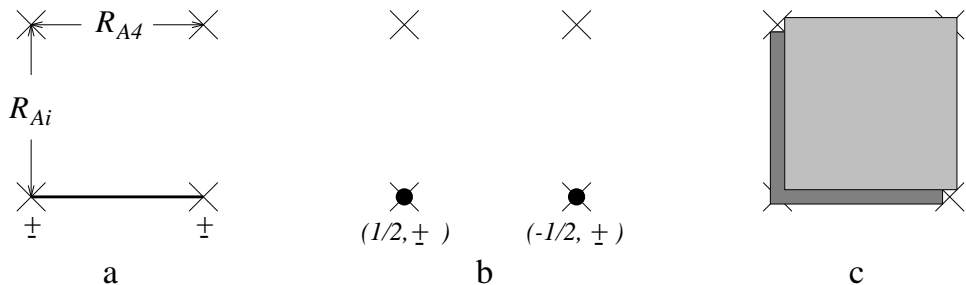}}
\caption{Non-BPS D-string (a), and its decay channels (b),(c).}
\end{figure}

Other decay channels become available to the D-string when the other
distances $R_{Ai}$ ($i=1,2,3$) become small. In particular, the
D-string along $x^4$ can decay into a pair of 
D2-branes carrying opposite bulk charges, {\it i.e.} a D2-brane and
an anti-D2-brane, and wrapping the $(x^i,x^4)$ cycle (Fig.~1c). 
Since the mass of each D2-brane in the orbifold metric is
$R_{Ai}R_{A4}/(2g_A)$, the D-string is stable in this channel when
\be
 R_{Ai} > {1 \over \sqrt{2}} \quad (i=1,2,3)\,.
\label{IIA_stab2}
\ee
The D-string can therefore also be thought of as a bound state of
two BPS D2-branes.
This decay channel can also be understood from the appearance of a
tachyon on the D-string carrying one unit of winding in the $x^i$
direction, when $R_{Ai}<1/\sqrt{2}$ \cite{Sen5}, 
or alternatively from the appearance of a tachyon between the
two D2-branes when $R_{Ai}>1/\sqrt{2}$.
In terms of the heterotic string, these decay channels are described
by (\ref{het_decay2}), and again the domains of stability are
qualitatively the same. There are analogous regimes of stability for
D-strings stretched between any two fixed points.

In the blow up of the orbifold to a smooth K3, the non-BPS D-strings
correspond to membranes wrapping pairs of shrinking 2-cycles. Since
such curves do not have holomorphic representatives, the states are
non-BPS. For each pair of 2-cycles there are four states, associated
with the different orientations of the membrane; the
membrane can wrap both cycles with the same orientation, or with
opposite orientation.  In either case the net bulk charge due to
$B=1/2$ can be made to vanish by turning on an appropriate
world-volume gauge field strength ($F=\pm 1$ in the first case, and
$F=0$ in the second).  The decay of the non-BPS D-string into a pair
of fractional BPS D-particles is described in this picture as the
decay of this membrane into two separate membranes, that wrap
individually around the two 2-cycles.

The entire discussion above also has a parallel in the T-dual
theory, type IIB on $T^4/\bbbz'_2$. The fractional BPS D-particles are
T-dual to BPS D-strings stretched between pairs of orbifold points,
and the non-BPS D-string we found is T-dual to the non-BPS D-particle
that was constructed in \cite{BG2}. As was demonstrated by Sen
\cite{Sen2}, this state can be obtained as a bound state of two
fractional BPS D-strings carrying opposite bulk charges, and
appropriate twisted charges. For sufficiently large $R_B$, the
D-string pair develops a tachyonic mode and decays into the non-BPS
D-particle.

\subsection{Non-BPS D-molecule}

The heterotic theory also contains states that are charged under
a single $U(1)$ associated with one fixed plane in the IIA orbifold,
but that are uncharged with respect to any other $U(1)$. The lightest
such states have $N_L=0$ and an internal weight vector of the form
\be
 V=\pm (0^{2n}, 2,\pm 2,0^{14-2n}) \,.
\label{het_nonbps_D0}
\ee
Since $V^2=8$, we must choose $N_R=c_R+3$ to satisfy level matching.
The state is therefore non-BPS\footnote{The degeneracy of the state
is rather large, and it actually contains $60$ long supermultiplets.} 
and its mass is given by
\be
 m_h = 2 \sqrt{6} \,.
\ee     
This state may decay into two BPS states of the form 
(\ref{D0weights}), or into two non-BPS states of the form 
(\ref{intern}). The latter possibility is energetically forbidden
since $2 \times 2 \sqrt{2} > 2 \sqrt{6}$, and the former is possible
provided that $R_{h4} > 1/(2\sqrt{6})$. In the heterotic theory, this
non-BPS state is therefore stable if \footnote{We are only
considering possible decay processes into states with trivial winding
number.}  
\be
R_{h4} < {1 \over 2 \sqrt{6}} \,.
\ee

The above suggests that the dual type IIA theory contains a bound
state of two fractional D-particles which are located at the same
fixed plane, and which carry opposite bulk charges. In the previous
subsection we saw that a bound state of two fractional D-particles of
opposite bulk charge that are located on {\em different} fixed planes
could be better described as a non-BPS D-string. Let us therefore
attempt to describe the above state as a non-BPS D-particle at a fixed
plane. The associated boundary state would then be given by 
\be
\label{D0p}
|D0_{nonbps};\pm> = c \Big(|U0>_{NSNS} 
\pm |T0>_{RR}\Big) \,,
\ee
where $|U0>_{NSNS}$ and $|T0>_{RR}$ are the same states as in
(\ref{D0}), and $c$ is a normalisation factor which will be determined
below. The resulting spectrum of open strings beginning and ending on
this D-particle is given by 
\be
[NS - R] \; c^2 \left(1 + (-1)^F \I_4 \right) \,.
\label{nonbps_D0_strings}
\ee
In order for this to make sense as the spectrum of an actual open
string theory, the normalisation should be $c=1/\sqrt{2}$.
On the other hand, the magnitude of the twisted charge in
(\ref{D0p}) would then be $\sqrt{2}$ in units of the twisted charge
associated to the fractional BPS D-particles (\ref{D0}).\footnote{
We thank A. Sen for pointing this out.} Furthermore, the spectrum of
open strings between the non-BPS D-particle and a fractional BPS
D-particle is the same as above, except that the overall factor is
$c/2$ rather than $c^2$. For this to make sense we need $c=1$, rather
than $c=1/\sqrt{2}$. 
With this normalisation, the charge (and mass) in (\ref{D0p}) is then
precisely twice the twisted charge of a fractional BPS D-particle, and
the state described by (\ref{D0p}) is not stable. This is consistent
with the fact that the open string spectrum is now doubled, and
therefore cannot describe an open string that begins and ends on 
a single D-particle. The situation is also different from the case of
the non-BPS D-string in that the pair of fractional BPS D-particles
that carry opposite bulk charges but equal twisted charges does not
exhibit a tachyonic instability, as follows from (\ref{frac_D0_open}). 

On the other hand, one should expect that two such fractional
D-particles can bind, since their interaction is of the form
$$
  V(r) = - {a\over r^7} + {b\over r^3} \;,\quad a,b>0\;,
$$
where the first term is the ten-dimensional (bulk) contribution,
and the second term is the six-dimensional (twisted) contribution.
Unlike the case of two BPS D-particles at different fixed planes,
this bound state does not correspond to a new D-brane;
it is most appropriately referred to as a `D-molecule'.
The D-molecule carries two units of twisted charge, but no 
bulk charge, and is therefore still restricted to the fixed plane.
(This is to be contrasted with the (threshold) bound state
of two fractional BPS D-particles carrying equal bulk charges and
opposite twisted charges, which corresponds to a bulk BPS 
D-particle). Since the above interaction comes from the one loop open
string diagram, it is ${\cal O}(g_A)$. At weak coupling the mass of 
the D-molecule is therefore well approximated by the mass of
the two fractional D-particles, {\it i.e.} $1/g_A$.

The decay channels described above for the heterotic string correspond
in IIA to the decay of the D-molecule into a pair of fractional BPS
D-particles or into a pair of non-BPS D-strings. Here it is the former
which is energetically forbidden.  On the other hand, the mass of two
non-BPS D-strings is $\sqrt{2} R_{Ai} / g_A$, and is therefore smaller
than that of the D-molecule if $R_{Ai}<1/\sqrt{2}$.  The D-molecule is
thus stable in the type IIA theory when $R_{Ai}>1/\sqrt{2}$.

Unlike the non-BPS D-string, the stability domains of the D-molecule
are qualitatively different in the two theories, {\it i.e.}
at weak and strong IIA coupling. 
At weak IIA coupling only the decay into non-BPS states is possible,
whereas at strong IIA coupling, only the decay into BPS states is
allowed. As the coupling is varied from weak to strong, the energy
levels must therefore cross over, and we expect that at intermediate
coupling, both decay channels will be available.

In the blow up to a smooth K3 the D-molecule corresponds
to {\it two} D2-branes wrapping a shrinking 2-cycle.
The world-volume gauge field must be $F=-1$ on
one of the membranes, to cancel the bulk charge due to $B=1/2$ on
the 2-cycle. The (conditional) stability of this particle implies 
that the two wrapped D2-branes should form a (non-threshold) bound 
state in a non-vanishing region of moduli space.

\subsection{Other non-BPS states}

\noindent There exist also other non-BPS states in the heterotic 
string that are stable in certain regions of the moduli space,
such as states transforming in the spinor representation
of $SO(32)$, {\it e.g.} $V=((1/2)^{16})$.
In $D=10$ this state has been identified with a $\bbbz_2$-valued
non-BPS D-particle in the dual type I string \cite{Sen3,Sen4,Witten2}.
Going through the sequence of duality transformations in
(\ref{sequence}) suggests the following interpretation for this
state: after the four T-dualities the D-particle becomes a non-BPS
D4-brane in IIB on $T^4/\bbbz'_2$. Under S-duality this transforms
into a non-BPS (non-Dirichlet) 4-brane. 
The final T-duality then gives a non-BPS
4-brane in IIA on $T^4/\bbbz_2$.
Like the type I D-particle, this 4-brane is
$\bbbz_2$-valued. Perhaps this 4-brane can be understood
as a bound state of an NS-5-brane and an anti-NS-5-brane,
in analogy with the D-brane case. However, it is not yet clear what
the analogue of the tachyon condensation would be.

\section{Conclusions}
\setcounter{equation}{0}

In this paper we have analysed the duality between the heterotic
string on $T^4$ and the type IIA string on K3 for states that are not
necessarily BPS. In particular, type IIA string theory on
$T^4/\bbbz_2$, which is the orbifold limit of K3, admits a non-BPS
D-string as well as a non-BPS D-molecule, which are related by the
duality map to perturbative non-BPS states in the heterotic string,
and therefore probe the duality beyond the regime of BPS states. The
D-string is also related by T-duality to the non-BPS D-particle of the
related orbifold of type IIB string theory, which was constructed in
\cite{BG2}.   

These states are not stable everywhere in moduli space.  We have
determined their regions of stability in both the heterotic and type
IIA pictures, and we have found these regions to be of non-vanishing
size in both cases. For the case of the non-BPS D-string, the regions
of stability are also qualitatively related by the duality map. Since
the masses of non-BPS states are not protected by supersymmetry, this
was not guaranteed a priori. 

It would be interesting to understand these branes in terms of the
K-theoretic framework proposed by Witten \cite{Witten2}.  More
generally, it would be interesting to analyse systematically the
various non-BPS Dirichlet-branes in orbifolds and orientifolds of type
II theories, and relate them to the K-theory predictions.

\section*{Acknowledgements}

We would like to thank Mina Aganagic, Eric Gimon, and Joe Minahan for
useful conversations, 
and Ashoke Sen for pointing out an error in an earlier version.  
O.B. is supported in part by the DOE under
grant no. DE-FG03-92-ER 40701.  M.R.G. is supported by a College
Lectureship of Fitzwilliam College, Cambridge.

\end{document}